\documentclass[aps,prl,twocolumn,superscriptaddress,showpacs]{revtex4-1}
\usepackage{graphicx}
\usepackage{amssymb}
\usepackage{amsmath}
\usepackage{appendix}
\usepackage{comment}

\begin{document}

\title{Enhanced Spin Conductance of a Thin-Film Insulating Antiferromagnet}

\author{Scott A. Bender}
\affiliation{Utrecht University, Princetonplein 5, 3584 CC Utrecht, The Netherlands}

\author{Hans Skarsv\aa g}
\affiliation{Department of Physics, Norwegian University of Science and Technology, NO-7491 Trondheim, Norway}

\author{Arne Brataas}
\affiliation{Department of Physics, Norwegian University of Science and Technology, NO-7491 Trondheim, Norway}

\author{Rembert A. Duine}
\affiliation{Utrecht University, Princetonplein 5, 3584 CC Utrecht, The Netherlands}
\affiliation{Department of Applied Physics, Eindhoven University of Technology,
P.O. Box 513, 5600 MB Eindhoven, The Netherlands}

\begin{abstract}
We investigate spin transport by thermally excited spin waves in an antiferromagnetic insulator. Starting from a stochastic Landau-Lifshitz-Gilbert phenomenology, we obtain the out-of-equilibrium spin-wave properties. In linear response to spin biasing and a temperature gradient, we compute the spin transport through a normal metal$|$antiferromagnet$|$normal metal heterostructure. We show that the spin conductance diverges as one approaches the spin-flop transition; this enhancement of the conductance should be readily observable by sweeping the magnetic field across the spin-flop transition. The results from such experiments may, on the one hand, enhance our understanding of spin transport near a phase transition, and on the other be useful for applications that require a large degree of tunability of spin currents.  In contrast, the spin Seebeck coefficient does not diverge at the spin-flop transition. Furthermore, the spin Seebeck coefficient is finite even at zero magnetic field, provided that the normal metal contacts break the symmetry between the antiferromagnetic sublattices.   

\end{abstract}

\maketitle

{\it Introduction}.~Antiferromagnets have recently garnered increasing interest in the spintronics community, both for their novel intrinsic properties and their technological potential.  Their appealing features are their lack of stray magnetic fields, fast dynamics relative to ferromagnets, and robustness against external fields \cite{Baltz:2016vb}. The last property is a double-edged sword, as the lack of response to an external field makes control of antiferromagnets challenging. Recent theoretical and experimental work has instead sought to generate and detect antiferromagnetic dynamics optically \cite{GomezAbal:2004il,*Kampfrath:2010kl,*Satoh:2010dr} and electrically \cite{Wadley:2016dv,*Hahn:ie,*Wang:2014fk,*Moriyama:2015fi,*Ross:2015ec,*Khymyn:2016cj}.


Spin transport through insulators is of particular interest since there is no dissipation associated with the  motion of electrons. However, there is currently a lack of understanding how spins can flow between metals via antiferromagnetic insulators. Exploring these phenomena is essential for exploiting antiferromagnetic insulators in a more active role in spintronics. In ferromagnets, equilibrium thermal fluctuations generate spin waves that can drive coherent magnetic dynamics \cite{Yan:2011he,*Tatara:2015hf} or transport spins \cite{Goennenwein:2015jn,*Cornelissen:2015cz,*Li:2016jf}.  For instance, a non-local spin conductance contains signatures of the spin transport properties. Measurements of this spin conductance have generated considerable excitement in the spintronics community \cite{Cornelissen:2015cz}. It is of interest to see if thermal magnons can provide a similar long-range spin-transport in $anti$ferromagnetic insulators. We predict that the spin conductance is as substantial in antiferromagnets and therefore expect that thermal magnon transport in these systems will generate a sizeable interest as well.  Below we discuss in detail two scenarios to open the door for long-range spin transport through antiferromagnetic insulators, without the need for adjacent ferromagnets, or, in principle, magnetic fields. 

In antiferromagnets, at zero magnetic fields, spin-wave excitations are doubly degenerate. The two branches carry opposite spin polarity.  Thus, to realize spin transport by thermally generated spin waves, the symmetry between the antiferromagnetic sublattices must be lifted.  One means of achieving this is to employ a ferromagnetic layer, controlled by a magnetic field \cite{Lin:2016cs}. Alternatively, the magnetic field itself suffices to break the sublattice symmetry, eliminating the need for a ferromagnetic component. Ref.\ \cite{Wu:2016hh,Seki:2015es} measured the spin Seebeck effect \cite{Ohnuma:2013il,Rezende:2016bv}, in which angular momentum is driven by a temperature gradient in bipartite electrically insulating antiferromagnets at finite magnetic fields.

The first scenario is the injection of thermal magnons by a spin accumulation in an adjacent metal. While spin accumulation-induced thermal magnon injection in ferromagnet$|$normal metal heterostructures has been the subject of recent theoretical research \cite{Cornelissen:2016ji,Zhang:2012hh}, predictions for the $anti$ferromagnetic analogue are currently lacking and are restricted to coherent magnetic dynamics of the antiferromagnetic order and the resulting spin superfluidity that requires external fields \cite{Takei:2014wd,Qaiumzadeh:2017jc}. Here, we show that the spin conductivity of thermal magnons is strongly enhanced upon approaching the spin-flop transition. This leads to a large amount of tunability of the magnon transport by an external field which may be desirable for applications.  

A second possibility for engineering magnon spin transport in antiferromagnets is to break the interface sublattice symmetry. Magnetically uncompensated antiferromagnet$|$metal interfaces have been studied theoretically \cite{Cheng:2014fj}. Nevertheless, the possibility of realizing a spin Seebeck effect by breaking the sublattice symmetry at the interface has not been proposed until now.

\begin{figure}[pt]
\includegraphics[width=\linewidth,clip=]{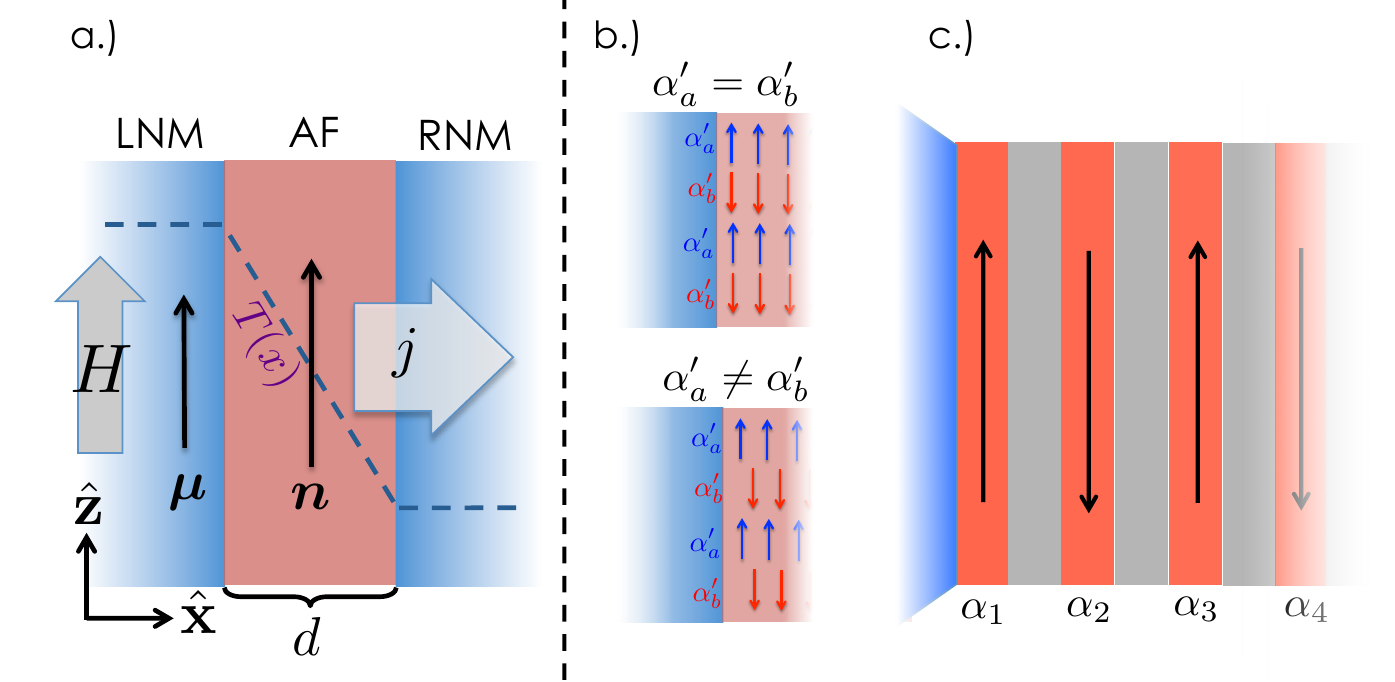}
\caption{a.) Left normal metal (LNM)/antiferromagnet (AF)/right normal metal (RNM) setup. A spin accumulation $\boldsymbol{\mu}=\mu\mathbf{z}$ at the left interface inside the left normal metal and temperature gradient $\partial_x T$ across the heterostructure are applied; as a result a spin current $j$ flows across the right interface. b.) Metal$|$antiferromagnet interface, with unbroken ($\alpha'_a=\alpha'_b$, top) and broken ($\alpha'_a\neq\alpha'_b$, bottom) sublattice symmetries. c.) Symmetry breaking in a synthetic antiferromagnet, composed of alternating metallic spacers and ferromagnetic layers with respective damping parameters $\alpha_1,\alpha_2\dots$.  If, for example, the contact (blue) material differs from that of the interlayer spacer (gray), then $\alpha_1 \neq \alpha_2$, analogously to b.)}.
\label{sch}
\end{figure}

{\it Stochastic Dynamics}. We consider a bipartite antiferromagnet (AF). The system is translationally invariant in the $yz$ plane. There is an interface along the plane $x=-d/2$ on the left with a normal metal (LNM) and an interface along the plane $x=d/2$ with an identical normal metal (RNM) on the right (see Fig.~\ref{sch}a). Let us suppose that a spin accumulation $\boldsymbol{\mu}=\mu \hat{\mathbf{z}}$ is fixed by, e.g., spin Hall physics in the left lead, or that a linear phonon temperature profile is established across the structure \footnote{A complete treatment of transport generally requires treating the coupled magnetic, phononic and electronic degrees of freedom of the heterostructure on equal footing.  However, if the metallic leads are good spin sinks (so that any spin accumulation driven by magnetic dynamics is quickly relaxed, and $\boldsymbol{\mu}$ is determined exclusively by spin Hall driving) and assuming a large phononic heat conductance throughout the structure, it is reasonable to neglect the feedback on the lead-electrons and phonons from the AF spin wave degrees of freedom in our simple model.}.

We begin by parameterizing the AF spin degrees of freedom in the long wavelength limit by the N$\acute{\rm{e}}$el order unit vector $\mathbf{n}$ and dimensionless magnetization $\mathbf{m}$. At zero temperature, the AF relaxes towards a ground state which is determined by the free energy $U$ \cite{Hals:2011fy}:
\begin{equation}
\label{U}
U=s\int_\mathcal{V} d^3 r \left(\frac{\mathbf{m}^2}{2\chi} +\frac{A}{2} \sum_{i=1}^3 \left(\partial_i \mathbf{n} \right)^2-\frac{1}{2} K n_z^2-\mathbf{H}\cdot\mathbf{m} \right) \, .
\end{equation}
Here, $s=s_a+s_b$ is the sum of the saturation spin densities of the $a$ and $b$ sublattices (in units of $\hbar$), $\mathcal{V}$ is the volume of the AF, $\chi$ is the susceptibility, $A$ is the N$\acute{\rm{e}}$el order exchange stiffness and $K(>0)$ is the uniaxial, easy-axis anisotropy.  The external magnetic field $\mathbf{H}$ is taken to be applied along the $z$ direction in order to preserve rotational symmetry around the $z$ axis in spin space. The bulk symmetry of the bipartite lattice under the interchange of the sublattices, which sends $\mathbf{m}\rightarrow \mathbf{m}$ and $\mathbf{n}\rightarrow {-\mathbf{n}}$, is manifest in the form of $U$. 

At sufficiently small magnetic fields, $\left| H \right|<H_c=\sqrt{K/\chi}$, the ground states are degenerate, given by $\mathbf{n}=\pm \mathbf{z}$ and $\mathbf{m}=0$, and the AF is in the antiferromagnetic phase.  In the antiferromagnetic phase, the ground state magnetic texture is insensitive to the spin accumulation $\boldsymbol{\mu}$ in the linear response, and the AF does not support a spin current at zero temperature.  At fields $\left| H\right|>H_c$, the ground state is ``spin-flopped", with $\mathbf{m}\propto{\mathbf{z}}$ and $\mathbf{n}$ in the $xy$ plane. Spin biasing of the spin-flopped state generates a spin super current \cite{Takei:2014wd,Qaiumzadeh:2016ti} at zero temperature.   In order to focus on transport by thermally activated spin waves, we restrict the following discussion to the antiferromagnetic phase. Furthermore, in this phase, the spin waves are circular and therefore simpler to analyze. 

At finite temperatures, fluctuations drive the AF texture away from the zero temperature configuration, necessitating equations of motion that incorporate bulk and boundary fluctuations and dissipation. The small amplitude excitations of the N$\acute{\rm{e}}$el order above the ground state $\mathbf{n}=-\mathbf{z}$ are $\delta \mathbf{n}$ described by the linearized equation of motion in the bulk ($-d/2<x<d/2$): 
\begin{equation}
(\partial_x^2+\mathfrak{q}^{2})n=-\mathfrak{f}_{B}/A\, .
\label{bulk}
\end{equation}
(see Supplemental Material). Here, $n=n(x,\mathbf{q},\omega)$ is the Fourier transform (in the coordinates $\boldsymbol{\rho}=y\hat{\mathbf{y}}+z\hat{\mathbf{z}}$ and $t$) of $n(x,\boldsymbol{\rho},t)\equiv n_x (x,\boldsymbol{\rho},t)+i n_y(x,\boldsymbol{\rho},t)$, while $\mathfrak{q}^{2}\equiv -\mathbf{q}^2-K/A+ \eta_{\omega}^2/A\chi+i\alpha \hbar \omega/A$ with $\eta_{\omega}\equiv \chi(\hbar \omega+H)$.  The stochastic force $\mathfrak{f}_{B }$, modeling fluctuations of the AF lattice that drive $n$, is connected to the bulk Gilbert damping $\alpha$ by the fluctuation dissipation theorem (here in the large exchange regime, $\chi^{-1}\gg \hbar \omega+H$):
\begin{align}
\nonumber
\langle \mathfrak{f}^*_{B} (x,\mathbf{q},\omega) \mathfrak{f}_{B} (x',\mathbf{q},\omega')\rangle = \\
\times \delta(x-x')\delta(\mathbf{q}-\mathbf{q}')\delta(\omega-\omega')\frac{2\alpha (2\pi)^3 \hbar \omega}{\rm{tanh} [\hbar \omega/2\it{T}]}\, ,
\label{bfdt}
\end{align}
where $T=T(x)$ is the local temperature in units of energy. 

Complementing Eq.~(\ref{bulk}) are boundary conditions on $n$:
\begin{align}
\nonumber
A\partial_x n+i \alpha_\omega' d(\hbar \omega-\mu) n=-\mathfrak{f}_{L}\, \,\,\,\, (x=-d/2)\\
-A\partial_x n+i \alpha_\omega' d\hbar \omega n=-\mathfrak{f}_{R}\, \,\,\,\, (x=d/2)\, ,
\label{rightbc}
\end{align}
where $\mathfrak{f}_{L}$ and $\mathfrak{f}_{R}$ correspond to fluctuations by lead electrons at the interfaces. The quantity $\alpha_\omega'\equiv \alpha'-\eta_\omega \tilde{\alpha}'$, describing dissipation of magnetic dynamics at the interfaces (which we have taken to be identical for simplicity), has contributions from both sublattice-symmetry-respecting ($\alpha'$) and -breaking ($\tilde{\alpha}'$) microscopics there.  For example, in a simple model in which fluctuation and dissipation torques for the two sublattices are treated independently (see Supplemental Material), one finds $\alpha'=(\alpha_a'+\alpha_b')/2$ and $\tilde{\alpha}'=(\alpha_a'-\alpha_b')/2$; here $\alpha_\zeta '=g_\zeta^{\uparrow \downarrow}/4\pi s d$ (with $g^{\uparrow \downarrow}_\zeta$ as the spin mixing conductance) is the effective damping due to spin pumping for sublattice $\zeta=a,b$ \cite{Cheng:2014fj,Takei:2014wd}(see Fig.~\ref{sch}b). Such a model corresponds to the continuum limit of a synthetic antiferromagnet (composed of ferromagnetic macrospins separated by normal metals) in which sublattice symmetry breaking may be more carefully controlled (see Fig.~\ref{sch}c). 

The effective surface forces $\mathfrak{f}_{L(R)}$ and damping coefficient $\alpha_\omega'$ are connected via the fluctuation-dissipations theorems for the $l=L,R$ interfaces:
\begin{align}
\nonumber
\langle \mathfrak{f}^*_{l} (\mathbf{q},\omega) \mathfrak{f}_{l' } (\mathbf{q},\omega')\rangle = \\
\times \delta_{ll'}\delta(\mathbf{q}-\mathbf{q}')\delta(\omega-\omega')\frac{2s\alpha_\omega' (2\pi)^3(\hbar \omega-\mu_{l})}{\rm{tanh} [(\hbar \omega-\mu_{l})/2\it{T_{l}}]}\, ,
\label{sfdt}
\end{align}
where we have retained terms up to first order in $\eta_\omega$. Here $T_{L}$ and $T_{R}$ are lead electronic temperatures, and, in our setup, $\mu_L=\mu$ and $\mu_R=0$.

{\it Spin Transport}.~We now obtain the spin current that flows across the right interface in linear response to the spin accumulation $\boldsymbol{\mu}=\mu\hat{\mathbf{z}}$ at the left interface. Rewriting the equation of motion for the magnetization $\mathbf{m}$ (Eq.~(2) in the supplementary material) as a continuity equation for the spin density $\mathbf{s}=s\hbar \mathbf{m}$, one obtains an expression for the spin current: $\mathbf{j}=-s A\mathbf{n}\times \partial_x\mathbf{n}$.  Solving Eqs.~(\ref{bulk})-(\ref{sfdt}) in the absence of a temperature gradient, retaining terms only up to linear order in $\mu$, the $z$-spin current flowing through the right interface becomes:
\begin{align}
j\equiv \langle \hat{\mathbf{z}}\cdot \mathbf{j} \rangle=As\mathrm{Im} \langle  n^*(\mathbf{r})\partial_x n(\mathbf{r})\rangle_{x=\frac{d}{2}}=G \mu\, ,
\end{align}
where we have introduced the spin conductance $G$. 

In the low-damping/thin-film limit, $d \ll \lambda$, where $\lambda^2\equiv A/\alpha T$ is the imaginary correction to $\mathfrak{q}^2$ (i.e. $\mathfrak{q}^2=q_r^2+i\lambda^{-2}$ with $q_r$ real) due to Gilbert damping, the spin current is carried by well defined spin-wave modes (corresponding to solutions to Eq.~(\ref{bulk}) in the absence of noise) with frequencies $ \omega_{l\mathbf{q}}^{(\pm)}=-H/\hbar\pm  \hbar^{-1}\sqrt{(A \mathbf{q}^2+A(l\pi/d)^2+K)/\chi}$. Here $\mathbf{q}$ is the transverse wavevector, $l$ is an integer denoting spin-wave confinement in the $x$ direction, and the labels $\pm$ corresponds to the two spin-wave branches for which $\mathbf{n}$ rotates in opposite directions, as $\omega^{(+)}_{l\mathbf{q}}$ has the opposite sign of $\omega^{(-)}_{l\mathbf{q}}$ (though a different magnitude when $H\neq 0$) In the low damping thin/film limit, the spin conductance $G$ is therefore a sum over contributions from each of these modes and can further be broken into ``symmetric" and ``antisymmetric" (under interchange of the sublattices) pieces:
\begin{equation}
G=\sum_{l=0, 1,2 \dots}\int \frac{d^2q}{(2\pi)^2}\left(G^{(\mathrm{S})}_{l\mathbf{q}}+G^{(\mathrm{A})}_{l\mathbf{q}}\right)\, .
\label{totG}
\end{equation}
 Defining:
\begin{equation}
F^{(i,j)}_\pm(\xi/T)\equiv \frac{  \left( T/\xi\right)^i\left(\hbar \omega^{(\pm)}/T\right)^{1+j}}{\mathrm{sinh}^2 \left(\hbar \omega^{(\pm)}/2T \right)}\, ,
\label{gs}
\end{equation}
with $\hbar \omega^{(\pm)}=-H\pm \xi$, the symmetric contribution, which is proportional to $\alpha'$, is:
\begin{equation}
G_{l\mathbf{q}}^{(\mathrm{S})} =\frac{( \varsigma_l \alpha')^2}{2 \varsigma_l \alpha'+\alpha} \left(\frac{1}{\chi T}\right) \mathcal{G}^{(\mathrm{S})}(\xi_{l\mathbf{q}}/T)
\label{gs0}\, ,
\end{equation}
where $\xi_{l\mathbf{q}}\equiv \sqrt{A(\mathbf{q^2}+[l\pi/d]^2)/\chi+K/\chi}$ and $\mathcal{G}^{(\mathrm{S})}(\xi/T)=F^{(1,0)}_+(\xi/T)-F^{(1,0)}_-(\xi/T)$, with $\varsigma_l=1$ for $l=0$ and $\varsigma_l=2$ for $l> 0$ (reflecting the exchange boundary conditions \cite{Kapelrud:2013ht,Hoffman:2013bl}). The antisymmetric piece, which is proportional to $ \tilde{\alpha}'$, reads: 
\begin{equation}
G_{l\mathbf{q}}^{(\mathrm{A})} =\varsigma_l \tilde{\alpha}'\frac{2 \varsigma_l  \alpha'  ( \varsigma_l \alpha'+\alpha)}{(2 \varsigma_l \alpha'+\alpha)^2}  \mathcal{G}^{(\mathrm{A})}(\xi_{l\mathbf{q}}/T)
\label{ga0}
\end{equation} 
where $\mathcal{G}^{(\mathrm{A})}(\xi/T)=F^{(0,0)}_+(\xi/T)+F^{(0,0)}_-(\xi/T)$ and we have assumed that $\tilde{\alpha}'\lesssim\alpha'$.  From Eqs.~(\ref{gs0}) and~(\ref{ga0}) we find an algebraic decay of the spin current with film thickness.  In the extreme thin film limit, $d\ll g^{\uparrow \downarrow} /4\pi s \alpha$, the damping at the interface dominates over bulk, and both contributions decay as $1/d$; in the opposite regime, $d\gg g^{\uparrow \downarrow} /4\pi s \alpha$, one has that both again decay in the same way, as $1/d^2$. The symmetric and antisymmetric contributions may instead be distinguished by reversing the direction of the applied field: $\mathcal{G}^{(\mathrm{A})}(\xi/T)$ changes sign under $H\rightarrow -H$ (and therefore vanishes at zero field), while $\mathcal{G}^{(\mathrm{S})}(\xi/T)$ remains the same.   Note however that the antisymmetric contribution, Eq.~(\ref{ga0}), is suppressed by a factor of $T \chi \ll 1$ relative to the symmetric contribution, Eq.~(\ref{gs0}).

\begin{figure}[pt]
\includegraphics[width=0.8\linewidth,clip=]{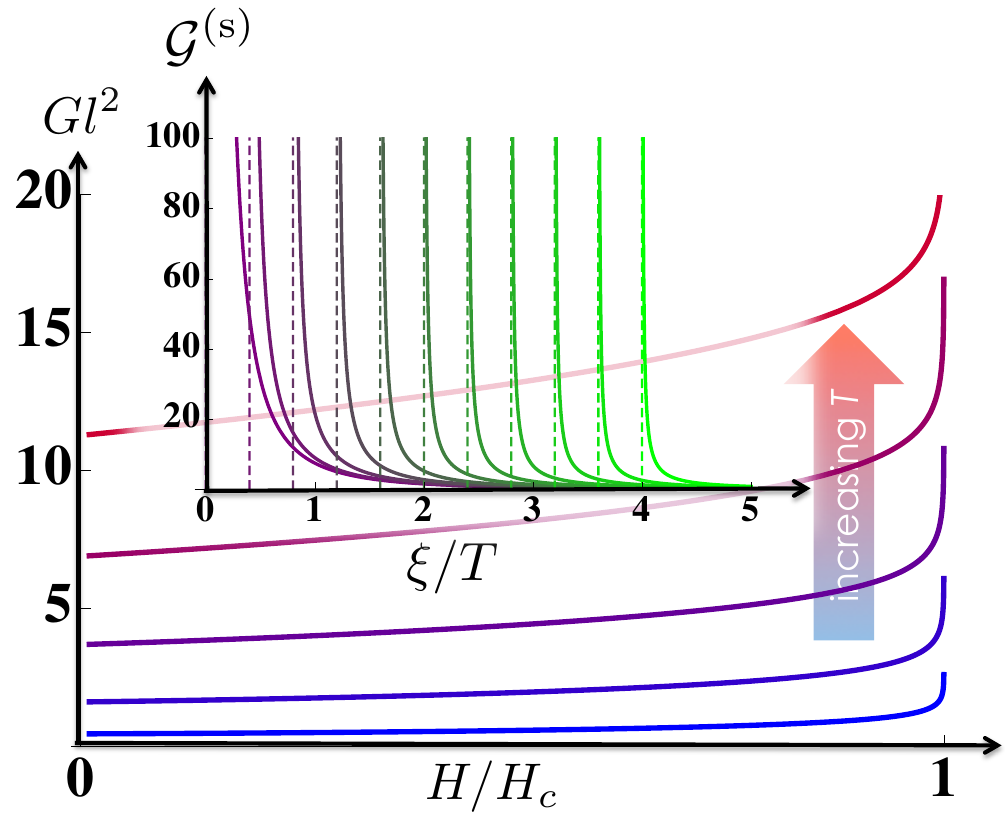}
\caption{Total conductance $G$ times $l^2\equiv A\chi$ for $\chi^{-1}=10H_c$, $d=l$, $\alpha=\alpha'=\tilde{\alpha}'=0.1$, for varying temperatures $T/H_c$ from 1 to 5 in steps of 1. Inset: symmetric magnon conductance $\mathcal{G}^{(\rm{S})}$ (defined below Eq.~(\ref{gs0})) for a single magnon mode with energy $\hbar \omega^{(+)}=-H+\xi$  as a function of $\xi$ for fixed temperature $T$. As $\xi \rightarrow H$, corresponding to the closing of the magnon gap ($H\rightarrow H_c$) and thus approaching the spin-flop transition, the conductance diverges. Shown are different fields (corresponding to the vertical dashed lines) for $0$ to $H_c$ in steps of $T/10$, with the corresponding curves for $\mathcal{G}^{(\rm{S})}$ shown in shades of purple to green (color online). $\mathcal{G}^{(\rm{A})}$ is qualitatively similar and therefore not shown.}
\label{gve}
\end{figure}

As a consequence of the divergence of the Bose-Einstein distribution at zero spin-wave gap (and as a precursor to superfluid transport \cite{Takei:2014wd}), the spin conductance diverges as one approaches the spin-flop transition. The boundary for the antiferromagnetic phase is defined by the vanishing of the spin-wave gap for one of the modes ($\hbar \omega_{0\mathbf{0}}^{(\pm)}=0$), which determines the critical field, $H_c=\sqrt{K/\chi}$. Then, from Eqs.~(\ref{gs0}) and~(\ref{ga0}), both the symmetric and antisymmetric contributions diverge as $1/(1-\left|H \right|/H_c)$ as $H\rightarrow H_c$ (see Fig.~\ref{gve}).  This enhancement of the spin-wave conductance is a key feature of spin transport in antiferromagetic insulators with a spin-flop transition. 

We may compare these results with spin transport driven by a temperature gradient. Supposing a linear temperature gradient $T(x)=T+(\partial_x T) x$, with a continuous profile across the structure so that $T_L=T-\partial_x T d/2$ and $T_R=T+\partial_x T d/2$, Eqs.~(\ref{bulk})-(\ref{sfdt}) yield a spin current for $\mu=0$: 
\begin{equation}
j=-S \Delta T\, ,
\end{equation}
where $\Delta T =d\partial_x T$ is the temperature change across the AF. In the low-damping/thin-film limit, the Seebeck coefficient $S$ similarly separates into symmetric and antisymmetric sums over discrete spin-wave modes:
\begin{equation}
\label{sshort}
S=\sum_{l=0, 1,2\dots}\int \frac{d^2 q}{(2\pi)^2}\left( S_{l\mathbf{q}}^{(\mathrm{S})} +S_{l\mathbf{q}}^{(\mathrm{A})}\right)\, .
\end{equation}
where
\begin{equation}
S_{l\mathbf{q}}^{(\mathrm{S})} =(\varsigma_l \alpha'/8) \left(1/ \chi T\right) \mathcal{S}^{(\mathrm{S})} \left(\xi_{l\mathbf{q}}/T\right)\, ,
\label{snqs}
\end{equation}
is the symmetric contribution, with $\mathcal{S}^{(\mathrm{S})} \left(\xi/T\right)\equiv F^{(1,1)}_+(\xi/T)-F^{(1,1)}_-(\xi/T)$, and 
\begin{equation}
S_{l\mathbf{q}}^{(\mathrm{A})} =\varsigma_l (\tilde{\alpha}'/4)  \mathcal{S}^{(\mathrm{a})} \left(\xi_{l\mathbf{q}}/T\right)\, ,
\label{snqa}
\end{equation}
the antisymmetric contribution, with $\mathcal{S}^{(\mathrm{A})} \left(\xi/T\right)\equiv F^{(0,1)}_+(\xi/T)+F^{(0,1)}_-(\xi/T)$. In contrast to the spin conductance, there is no divergence in the spin Seebeck coefficient as $H\rightarrow H_c$.  Furthermore, the antisymmetric contribution is $even$ under $H\rightarrow -H$ (and is generally nonzero at zero field), while the symmetric contribution is $odd$ (vanishing at zero field, as is required by sublattice symmetry); in contrast to spin biasing, a temperature gradient requires either a field or sublattice symmetry breaking at the interfaces in order to generate a spin current, else the two branches $\pm$ carry equal and opposite spin currents. Both symmetric and antisymmetric contributions to $S$ decay as $1/d$; writing $j=S\Delta T=\varsigma \partial_x T $, one finds that $\varsigma\equiv Sd$ is constant, reflecting that the Seebeck effect here is driven by bulk fluctuations.

\begin{figure}[pt]
\includegraphics[width=0.8\linewidth,clip=]{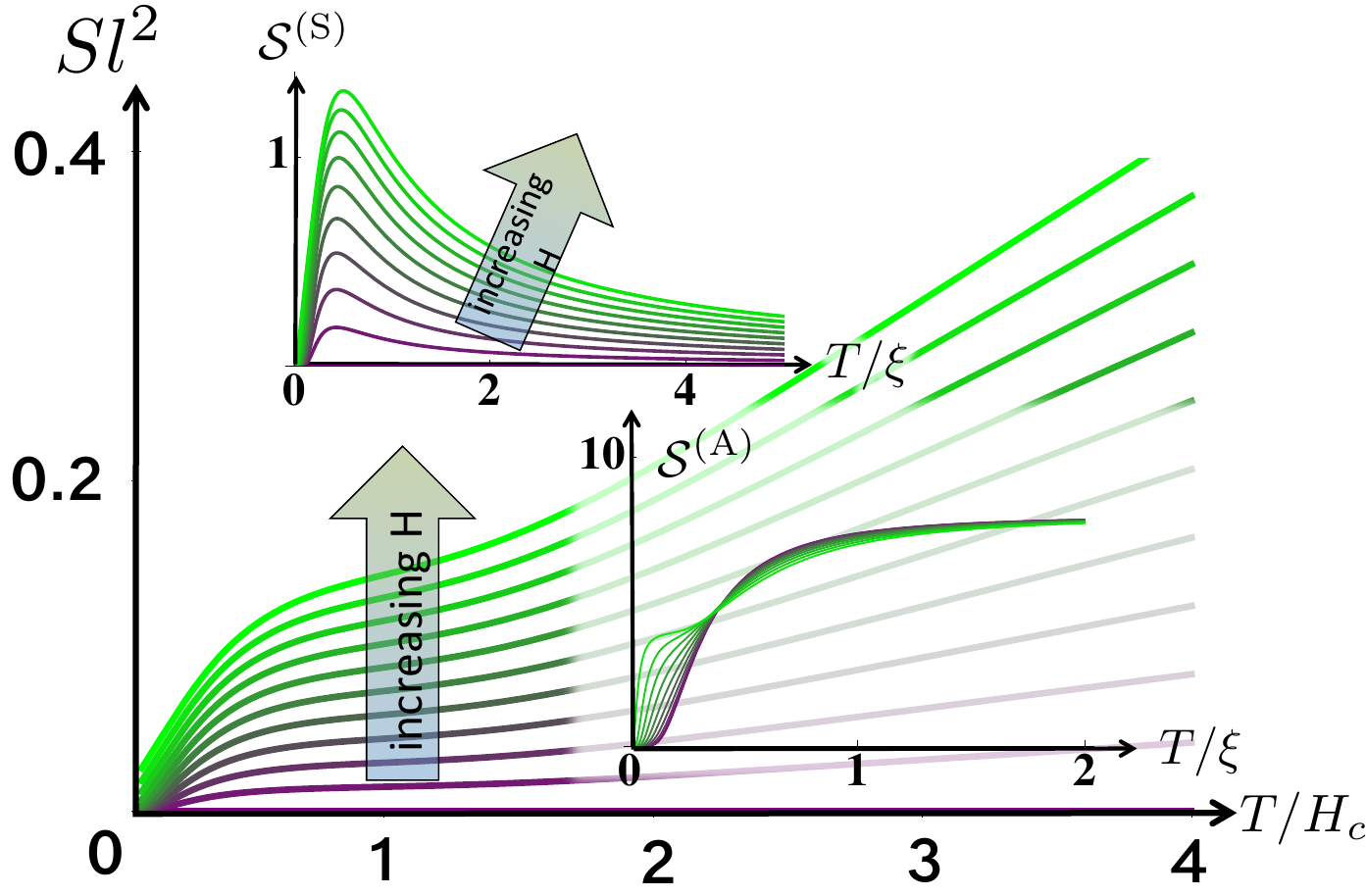}
\caption{Main figure: total spin Seebeck coefficient $S$ for the symmetric case ($\tilde{\alpha}'=0$), Eq.~(\ref{sshort}) for $\alpha'=0.1$, $d=l=\sqrt{A\chi}$, and $\chi^{-1}=10H_c$.  $\mathcal{S}^{(\rm{S})}$ (upper inset) and $\mathcal{S}^{(\rm{A})}$ (lower inset) as functions of temperature for fixed $\xi$.  While the symmetric contribution vanishes at high temperatures, the antisymmetric contribution saturates at $\mathcal{S}^{(\mathrm{A})}=8$. In all figures, the solid purple/green curves correspond to $H$ ranging from $0$ to $H_c$ in increments of $H_c/10$ (color online). At high temperatures, $S$ grows larger with temperature, in contrast with \cite{Wu:2016hh,*Seki:2015es} and \cite{Rezende:2016bv}; see second section of Supplemental Material.  }
\label{svt}
\end{figure}

 The transport coefficients $G$ and $S$ may be inferred from a number of different experiments. Suppose, for example, the leads are heavy metal with large spin-orbit interactions. The spin conductance $G$ may then be measured electrically as follows. Via the spin Hall effect, a spin accumulation is created in the LNM from an applied electric current $I_L$ flowing in the $y$ direction. In turn, the spin accumulation excites spin-waves in the AF, as described above. Via the inverse spin Hall effect, this spin current is then converted to a measurable electrical current flowing in the $y$-direction inside the RNM.  Under closed-circuit conditions, this current manifests as a voltage build-up $V_R$. For a platinum$|$MnF$_2|$platinum structure, we find a nonlocal resistance $R=V_R/I_L \sim m\Omega $ from our theory, which is of the same order of magnitude as that measured in \cite{Cornelissen:2016ji} for a ferromagnet.  Similarly, for the same setup, the Seebeck coefficient $S$ can be obtained by applying a temperature difference across the structure. We estimate that a temperature difference of $\Delta T=10^{-3} $K applied across a 10nm thick AF results in a voltage $\sim \mu$V, which is in the range of that measured by \cite{Wu:2016hh}. (See third section of Supplemental Material).  In addition to MnF$_2$, other materials, such NiO \cite{Prakash:2016ed} and Cr$_2$O$_3$ \cite{Wang:2015hh,Baltz:2016vb}, are also possible condidates for the insulating antiferromagnetic layer.

 {\it Conclusion and Discussion.} In this Letter, we have theoretically demonstrated two methods to realize spin transport in thin antiferromagnetic insulators that do not require the presence of a magnetic field or a ferromagnet. Working from a stochastic Landau-Lifshitz-Gilbert phenomenology, we obtained two key results.  First, the spin conductance diverges as the magnetic field approaches the spin-flop transition.  Second, the spin Seebeck effect may survive at zero field if the symmetry between antiferromagnetic sublattices is broken at the interface with normal metal contacts. Additionally, we estimated the inverse spin Hall voltages that would be produced by spin and temperature biasing in experiments.

The thin-film approximation,~Eqs.~(\ref{totG}) to~(\ref{snqa}), in which the structural transport coefficients consist of contributions from well-defined spin-wave modes, are valid for thicknesses $d\ll \lambda$. The parameter $\lambda=\sqrt{A/\alpha T}$, which describes the decay of magnons across the thickness of the film, can be estimated from a Heisenberg model on a lattice as $\lambda\sim a \sqrt{T_N/T \alpha}$ (supposing $T\ll H,K$), where $T_N$ is the N$\acute{\rm{e}}$el temperature and $a$ the lattice spacing. For $T\sim T_N/10$, a low damping factor $\alpha\sim 10^{-3}$ and a lattice spacing $a\sim $nm, for example, $\lambda$ corresponds to a thickness of $\sim$50 nm, which grows larger at lower temperatures.

The stochastic Landau-Lifshitz-Gilbert phenomenology we employed, Eqs.~(\ref{bulk})-(\ref{sfdt}), may of course be extended to thicker films, resulting in, for example, an exponential decay over the lengthscale $\lambda^2 \left| k_T \right| \sim \sqrt{A \chi}/\alpha$ (with $\left| k_T \right|$ as the magnon thermal wavevector) rather than an algebraic decay of the spin conductance with distance. Thicker films, however, introduce additional complications, e.g., elastic disorder scattering and phonon-magnon coupling (e.g. phonon drag). Spin wave interactions (scattering and mean field effects), which are absent in the single particle treatment above, may change transport at higher spin wave densities, e.g. at higher temperatures/thicker films or near the spin-flop transition, where the Bose-Einstein divergence may necessitate a many-body treatment  thereby altering the conductance $G$.  The scattering times and length scales over which such effects become important remains an open question. In addition, in our model the transition from antiferromagnetic to spin-flop phase is second order; the presence of Dzyaloshinskii-Moriya interaction, spin wave interactions, or in-plane anisotropy can change critical exponents such as $\nu$ in $G\sim (1-\left|H \right|/H_c)^\nu$ from its value $\nu=-1$ obtained above or even alter order of the phase transition thereby modifying the field dependence of $G$ \cite{Hohenberg:1977vo}.

This work had received funding from the Stichting voor Fundamenteel Onderzoek der Materie (FOM) and the European Research Council via Advanced Grant number 669442 ``Insulatronics''.



%

\begin{widetext}

\section{Derivation of Spin Wave Equations of Motion} 

In this section, we obtain Eqs.~(2)-(5) from the free energy, Eq.~(1). We begin with the nonlinear coupled equations  for the N$\acute{\rm{e}}$el order and magnetization:
\begin{equation}
\label{ndot}
\hbar \dot{\mathbf{n}}=\mathbf{F}_m\times\mathbf{n}+\boldsymbol{\tau}_n\, ,
\end{equation}
\begin{equation}
\label{mdot}
\hbar \dot{\mathbf{m}}=\mathbf{F}_m\times\mathbf{m}+\mathbf{F}_n\times\mathbf{n}+\boldsymbol{\tau}_m\, .
\end{equation}
Here, $\mathbf{F}_m=-s^{-1}\delta U/\delta \mathbf{m}$ and $\mathbf{F}_n=-s^{-1}\delta U/\delta \mathbf{n}$, with $\delta$ representing a functional derivative  [17]. The terms $\boldsymbol{\tau}_n$ and $\boldsymbol{\tau}_m$ capture thermal fluctuations and dissipation in the bulk and at the interfaces. It is possible to find phenomenological expressions for  $\boldsymbol{\tau}_n$ and $\boldsymbol{\tau}_m$ by listing out all terms with the appropriate symmetries.  We take an alternative approach and construct $\boldsymbol{\tau}_n$ and $\boldsymbol{\tau}_m$ from the corresponding torques that would arise on two separate ferromagnetic sublattices [20]  (see Fig.~1b).  Momentarily neglecting the spin accumulation inside the left normal metal, the fluctuating and dissipative torques are:
\begin{equation}
\label{nzeta}
\left. \hbar \dot{\mathbf{m}}_\zeta \right|_{\rm{fd}}=\boldsymbol{\tau}_\zeta=\left[\mathbf{f}_\zeta -\alpha_\zeta \hbar \dot{\mathbf{m}}_\zeta\right]\times\mathbf{m}_\zeta\, ,
\end{equation}
where $\mathbf{m}_\zeta$ is a unit vector in the direction of the magnetization of the $\zeta=a,b$ sublattice.  The parameter $\alpha_\zeta$ depends on the coordinate $x$:
\begin{equation}
\alpha_\zeta(x)=\alpha+\alpha_\zeta 'd \delta (x+d/2)+\alpha_\zeta 'd \delta (x-d/2)\, ,
\end{equation}
where $\alpha$ is the bulk damping, while $\alpha_\zeta '=g_\zeta^{\uparrow \downarrow}/4\pi s d$, with $g^{\uparrow \downarrow}_\zeta$ as the spin mixing conductance, is the effective damping due to spin pumping [12,14], which may differ for the two sublattices (see Fig.~1c). (For simplicity, we assume that $\alpha'_\zeta$ is the same at the left and right interfaces.)  Meanwhile, the fluctuating forces $\mathbf{f}_\zeta$ have contributions from both the interfaces and the bulk:
\begin{equation}
\mathbf{f}_\zeta(\mathbf{r})=\mathbf{f}_{L \zeta}(\boldsymbol{\rho})\delta(x+d/2)+\mathbf{f}_{R \zeta}(\boldsymbol{\rho})\delta(x-d/2)+\mathbf{f}_{B\zeta}(\mathbf{r})\, ,
\end{equation}
where $\boldsymbol{\rho}=y\hat{\mathbf{y}}+z\hat{\mathbf{z}}$. The bulk and interface Langevin sources are subject to the fluctuation-dissipation relations:
\begin{align}
\label{cbfdt}
\langle f_{B \zeta}^{(i)}(\mathbf{r},t) f_{B \zeta '}^{(i')} (\mathbf{r}',t') \rangle=\alpha \delta_{\zeta \zeta'}\delta_{ii'}\delta(\mathbf{r}-\mathbf{r}')  R(x,t-t') \\
\langle f_{l \zeta}^{(i)}(\boldsymbol{\rho},t) f_{l' \zeta '}^{(i')} (\boldsymbol{\rho}',t') \rangle=\alpha_\zeta \delta_{ll'}\delta_{\zeta \zeta'}\delta_{ii'}\delta(\boldsymbol{\rho}-\boldsymbol{\rho}') R_l(t-t')\, ,
\label{csfdt}
\end{align}
with $l=L,R$.  The bulk and interface noise functions $R(x,t-t')$ and $R_{L(R)}(t-t')$ respectively depend on the bulk and left (right) interface temperatures $T(x)$ and $T_{L(R)}$. In the white noise limit, these are proportional to $\delta (t-t')$; we will consider colored noise, and because we will require only the Fourier transforms of the these quantities (see Eqs.~(3) and~(5)), we do not specify the time-dependence of the colored noise functions here. 

Then,
\begin{align}
\boldsymbol{\tau}_m=\frac{1}{2}(\boldsymbol{\tau}_a+\boldsymbol{\tau}_b)=\left(\mathbf{f}_m-\alpha_{S} \hbar \dot{\mathbf{m}} -\alpha_A \hbar \dot{\mathbf{n}}\right)\times \mathbf{m} 
+\left(\mathbf{f}_n-\alpha_{A} \hbar \dot{\mathbf{m}} -\alpha_S \hbar \dot{\mathbf{n}}\right)\times \mathbf{n}\, ,
\label{taum}
\end{align} 
while 
\begin{equation}
\boldsymbol{\tau}_n=\frac{\hat{P}}{2}\left(\boldsymbol{\tau}_a-\boldsymbol{\tau}_b \right)=\left(\mathbf{f}_m-\alpha_S\hbar \dot{\mathbf{m}}\right) \times\mathbf{n}
- \alpha_A\left(\hbar \mathbf{\dot{n}}\times\mathbf{n}+\hat{P} \hbar\dot{\mathbf{m}}\times \mathbf{m}\right) \, .
\label{taun}
\end{equation}
Here, $\hat{P}=1-\mathbf{n}(\mathbf{n}\cdot))$ projects out components colinear with $\mathbf{n}$, ensuring that $\left| \mathbf{n}\right|=1$, while $\mathbf{f}_m=(\mathbf{f}_a+\mathbf{f}_b)/2$, $\mathbf{f}_n=(\mathbf{f}_a-\mathbf{f}_b)/2$, $\alpha_S(x)=(\alpha_a(x)+\alpha_b(x))/2$, and $\alpha_A(x)=(\alpha_a(x)-\alpha_b(x))/2$. 

Finally, to include a spin accumulation $\boldsymbol{\mu}$ along the left interface, we replace $\hbar \dot{\mathbf{n}}_\zeta\rightarrow \hbar \dot{\mathbf{m}}_\zeta-\boldsymbol{\mu}\times \mathbf{m}_\zeta$ in the terms with $\delta (x+d/2)$ in Eq.~(\ref{nzeta}) above of the supplemental material (SM); correspondingly, $\hbar \dot{\mathbf{n}}\rightarrow \hbar \dot{\mathbf{n}}-\boldsymbol{\mu}\times \mathbf{n}$ and $\hbar \dot{\mathbf{m}}\rightarrow \hbar \dot{\mathbf{m}}-\boldsymbol{\mu}\times \mathbf{m}$ in the terms with $\delta(x+d/2)$ in SM Eqs.~(\ref{taum}) and~(\ref{taun}), above.  Inserting the expressions for $\boldsymbol{\tau}_n$ and $\boldsymbol{\tau}_m$ of SM Eqs.~(\ref{taum}) and (\ref{taun}) into SM Eqs.~(\ref{ndot}) and~(\ref{mdot}), we obtain the full nonlinear equations for the noisy antiferromagnetic dynamics.  Note that the terms proportional to $ \alpha_A$ in SM Eqs.~(\ref{taum}) and~(\ref{taun}) break the symmetry of SM Eqs.~(\ref{ndot}) and~(\ref{mdot}) under $\mathbf{n}\rightarrow -\mathbf{n}$, $\mathbf{m}\rightarrow \mathbf{m}$, reflecting the broken $a\leftrightarrow b$ sublattice symmetry at the interfaces.

Next, we expand SM Eqs.~(\ref{ndot}) and~(\ref{mdot}) around the ground state, which we have chosen, without loss of generality, as $\mathbf{n}=+\hat{\mathbf{z}}$. Writing $\boldsymbol{\mu}=\mu \hat{\mathbf{z}}$ in order to preserve rotational symmetry of the spin around the $z$ axis, we define for $\mathbf{a}=\mathbf{n}$, $\mathbf{m}$, $\mathbf{f}_m$ and $\mathbf{f}_n$ the Fourier transform in a circular basis:
\begin{equation}
a(x,\boldsymbol{q},\omega)=\int \frac{d^2\rho}{(2\pi)^2} \frac{dt}{2\pi}e^{i\omega t-i\mathbf{q}\cdot\boldsymbol{\rho}}
\left[ a_x(\mathbf{r},t)+ i a_y(\mathbf{r},t) \right]\, ,
\end{equation}
so the linearized equations of motion become: 
\begin{align}
\hbar \omega n=- \chi^{-1} m-H  n+ f_{m}+i \hbar \omega( \alpha_A n+\alpha_S m)
\label{nomega}
\end{align}
\begin{align}
\hbar \omega m=(A\partial_x^2-A\mathbf{q}^2-K)n- H m+ f_{n}
+i\hbar \omega( \alpha_S n+\alpha_A m)\, .
\label{momega}
\end{align}
Finally, we solve SM Eq.~(\ref{nomega}) for $m$, and insert the result into SM Eq.~(\ref{momega}) to obtain a differential equation for $n$.  Neglecting terms $\alpha^2$, $f \alpha$, etc., and supposing the strong exchange limit, $\chi^{-1} \gg \alpha_A \hbar \omega $ \footnote{Strictly speaking, this is not possible at the interfaces where $\alpha_A (x)=(\delta(x+d/2)+\delta(x-d/2))(\alpha_a-\alpha_b)/2$ is nonzero.  However, in order parameterize the AF by smoothly varying fields $\mathbf{m}$ and $\mathbf{n}$, it is necessary that $\chi^{-1}$ is large, so we may consider $\alpha_a(x)$ as a smoothly varying function which is sharply peaked at $x=\pm d /2$ with $\chi^{-1}\gg \hbar \omega \rm{Max}(\alpha_A)$; then Eq.~(\ref{totaln}) follows.}, one obtains (again omitting dependence on $\mu$, which can be easily restored as above, for brevity):
\begin{align}
\nonumber
0=(A\partial_x^2-K)n -\chi \left(\hbar \omega+H \right)i2 \tilde{\alpha}' d \sum_{l=L,R}(\hbar \omega-\mu_l)\delta(x-x_l)  \hbar \omega +\chi  \left(\hbar \omega+H \right)^2  n\\
+i \alpha \hbar \omega n+f_{n}-\chi \left( \hbar \omega+H\right)f_{m}+i \alpha' \hbar d \sum_{l=L,R}(\hbar \omega-\mu_l) \delta(x-x_l)\, ,
\label{totaln}
\end{align}
with $x_L=d/2$ and $x_R=-d/2$. In the bulk ($-d/2<x<d/2$), this yields Eq. (2) in the main text, with $\mathfrak{f}_B\equiv f_n-\eta_\omega f_m$ and $\eta_\omega=\chi(\hbar \omega+H)$. Neglecting terms $\sim \alpha \eta_\omega^2$ and using SM Eq.~(\ref{cbfdt}), one obtains, after Fourier transforming, Eq.~(3) of the main text. Integrating SM Eq.~(\ref{totaln}) over the interfaces and restoring dependence on the spin accumulations at the boundaries, one obtains the boundary conditions, Eqs.~(4) in the main text, where $\mathfrak{f}_{l}\equiv f_{ln}-\eta_\omega f_{lm}$ for $l=L,R$ and $\alpha_\omega'\equiv\alpha'-2 \eta_{\omega} \tilde{\alpha}'$, with $\alpha'\equiv(\alpha_a'+\alpha_b')/2$ and $\tilde{\alpha}'\equiv(\alpha_a'-\alpha_b')/2$.  From SM Eq.~(\ref{csfdt}), one obtains the surface fluctuation-dissipation theorems Eq.~(5) in the main text, in the large exchange limit, $\eta_\omega \ll  1$. 

\section{High temperature limit of transport coefficients} 

In the high temperature limit ($T\gg \hbar \omega$), both the symmetric and antisymmetric contributions to the spin conductance increase with temperature, as more thermally occupied modes are available for transport: $G_{l\mathbf{q}}^{\mathrm{(S)}}\sim T^2/(\xi_{l\mathbf{q}}^2-H^2)$ and $G_{l\mathbf{q}}^{\mathrm{(A)}}\sim HT/(\xi_{l\mathbf{q}}^2-H^2)$. In contrast, the antisymmetric Seebeck coefficient $S^{(\mathrm{A})}$ saturates at a constant value, while the symmetric contribution $S^{(\mathrm{S})}$ vanishes as $H/T$. The $total$ Seebeck coefficient $S$ in Eq.~(12) of the main text, however, ultimately increases with temperature at high temperatures (see Fig. 3 of the main text), in disagreement with the prediction of [9] and measurements of [7], both of which show a Seebeck signal vanishing at high temperatures.  This is a consequence of the long wavelength nature of our continuum Landau-Lifshitz-Gilbert treatment, which fails when the spin wave wavelength becomes comparable to the lattice spacing, i.e. when $T$ becomes of the order of the N$\acute{\rm{e}}$el temperature $T_N$, provided that $T_N\gg H_c$ (which is the case for MnF$_2$, for example, where $T_N\sim 5H_c$[20]).  This can be remedied by, e.g., artificially introducing momentum cutoffs or employing a lattice model.  Additionally, the temperature at which our stochastic theory deviates from measurements may be affected by scattering processes absent in our approach, which require a more sophisticated treatment.

\section{experimental estimates}

In this section we estimate the strength of the experimental signals expected in the spin Hall experiments discussed in the main text. 

First, consider the proposal for the nonlocal conductance measurement of $G$.  Here, an applied current $I_L$, corresponding to a current density $\mathbf{j}_L=j_L \hat{\mathbf{y}}$, is applied in the LNM. As described in the main text, this signal is converted into an electrical current density $\mathbf{j}_R=j_R \hat{\mathbf{y}}$ near the AF$|$RNM interface. Neglecting losses of the spin current due to Gilbert damping in the AF and averaging $j_R$ over the $xz$ cross-section of the RNM, one finds that $j_R/ j_L \approx (\lambda_{N}/2d_{N})\theta^2 G/(G+G_N)$. The subscript ``$N$" here denotes properties of the normal metal. Specifically, $\lambda_N$ is the spin-flip length, $d_N$ the thickness in the $x$ direction, $G_N=\hbar \sigma /(2e)^2\lambda_N$ is the spin conductivity (with $\sigma$ as the electrical conductivity), and $\theta$ the spin Hall angle of the normal metal leads.  We have assumed that $\lambda_N\ll d_N$.  As $H$ approaches $H_c$ so that $G$ diverges, $j_R/j_L\approx(\lambda_{N}/2d_{N})\theta^2 $ no longer depends on the properties of the AF. Then, $j_R/j_L\approx(\lambda_{N}/2d_{N})\theta^2 $ does not depend on $H_c$ and only weakly depends on temperature through $\lambda_N$ and $\theta$.

At lower fields, however, $j_R/j_L$ depends strongly on both the magnetic field and temperature through $G$. Let us take $d_N=100$nm-thick Pt for both normal leads (with $\lambda_N\approx 14$nm, $\sigma\approx 0.08/\mu\Omega $cm, and $\theta\sim 0.1$ [21]), and $d=10$nm-thick MnF$_2$ for the AF (with $\sqrt{A/\chi a^2}\sim T_N\sim 67 K$ and $H_c \sim 12 K$ [20], with lattice spacing $a\sim $\AA).  For the interface, we take $g_a^{\uparrow \downarrow}\sim g_b^{\uparrow \downarrow}\sim 1/\rm{nm}^2$, resulting in $\alpha'=(g_a^{\uparrow\downarrow}+g_b^{\uparrow\downarrow})/8\pi s d\sim 10^{-3}$; taking $l\sim $\AA\ one has $G \approx 50/\mathrm{nm}^2 \sim 10 G_N$ at $T=2 H_c$ at $H=0$.  One then finds the same order of magnitude as above,  $j_R/j_L \sim (\lambda_{N}/d_{N})\theta^2 \sim 10^{-3}$, which is limited primarily by the spin Hall angle and thickness of Pt; lowering the temperature to $T=H_c/2$, however, reduces $G$ by a factor of $\sim 50$ and $j_R/j_L$ by almost an order of magnitude. In short, the independence of $j_R/j_L$ of field and temperature is a signature of the enhancement of the spin conductance $G$ near the spin-flop transition.  

We can compare with similar experiments in $ferromagnets$, such as [10], which measures the nonlocal resistance, $R_{nl}=(j_R/j_L) L_R/\sigma$, with $L_R$ as the lead length. These measurements were performed with magnetic films down to thickness $d=200$nm;employing our stochastic theory to this thickness yields $G\approx 12/$nm$^2$, and thus (for $L_R=100 \mu$m) yields $R_{nl}\sim m\Omega$, which is of the same order of magnitude as that measured in [10].  

Second, in the same setup, the coefficient $S$ can be measured similarly.  Upon the application of a temperature difference $\Delta T$ across the AF, a spin current flows from the AF into the RNM. This pure spin current is then converted into an inverse spin Hall electrical current as above. One obtains a cross-section averaged current density $j_R=(\lambda_N/d_N)(2e/\hbar)\theta S \Delta T$. For $\Delta T=10^{-3} $K with $T=2H_c$ and, taking e.g. $\tilde{\alpha}'\sim \alpha'/10$ (so $S\approx 0.4/\rm{nm}^2$), this yields $j_R\sim10^7 \rm{A}/\rm{m}^2$. Under closed circuit conditions for a $100 \mu$m long lead, this translates into a voltage $\sim \mu$V, which is in the range of that measured by [7].

\end{widetext}
\clearpage

\end{document}